\documentclass[conference]{IEEEtran}
\ifCLASSINFOpdf
   \usepackage[pdftex]{graphicx}
   \DeclareGraphicsExtensions{.pdf,.jpeg,.png}
\else
\fi
\usepackage[T1]{fontenc}
\usepackage[latin1]{inputenc}
\usepackage[english]{babel}

\usepackage{multirow}
\usepackage{tabularx,calc}
\usepackage{color}
\usepackage{colortbl}
\usepackage[disable]{todonotes}

\usepackage{tikz,amsmath, amssymb,bm,color}
\usepackage{pgfplots}
\usepackage{environ}
\usepackage{ifthen}
\makeatletter
\newsavebox{\measure@tikzpicture}
\NewEnviron{scaletikzpicturetowidth}[1]{%
	\def\tikz@width{#1}%
	\begin{lrbox}{\measure@tikzpicture}%
		\BODY
	\end{lrbox}%
	\pgfmathparse{#1/\wd\measure@tikzpicture}%
	\BODY
}
\makeatother
\newboolean{compileforpublish}
\setboolean{compileforpublish}{true}
\newboolean{isaccepted}
\setboolean{isaccepted}{true}
\ifthenelse{\boolean{isaccepted}}
{
	\newcommand\copyrighttext{%
		\footnotesize \parbox[t]{.11\textwidth}{\copyright{} \the\year~IEEE.} \parbox[t]{.89\textwidth}{Personal use of this material is permitted. Permission from IEEE must be obtained for all other uses, in any current or future media, including reprinting/republishing this material for advertising or promotional purposes, creating new collective works, for resale or redistribution to servers or lists, or reuse of any copyrighted component of this work in other works.}}
}
{
	\newcommand\copyrighttext{%
		\footnotesize \centering This work has been submitted to the IEEE for possible publication.\\ Copyright may be transferred without notice, after which this version may no longer be accessible.}
}

\newcommand\copyrightnotice{%
	\ifthenelse{\boolean{compileforpublish}}
	{
		\begin{tikzpicture}[remember picture,overlay]
		\node[anchor=south,yshift=10.5pt] at (current page.south) {\parbox{\dimexpr\textwidth-\fboxsep-\fboxrule\relax}{\copyrighttext}};
		\end{tikzpicture}%
	}
}

\renewcommand{\IEEEtitletopspaceextra}{20pt}

\begin{document}
\title{Scenarios for Development, Test and Validation\\ of Automated Vehicles}

\author{\IEEEauthorblockN{Till Menzel, Gerrit Bagschik and Markus Maurer}
\IEEEauthorblockA{Institute of Control Engineering\\
Technische Universit\"at Braunschweig\\
Braunschweig, Germany\\
Email: \{menzel, bagschik, maurer\}@ifr.ing.tu-bs.de}
}

\maketitle%
\copyrightnotice%

\begin{abstract}
The latest version of the ISO~26262 standard from 2016 represents the state of the art for a safety-guided development of safety-critical electric/electronic vehicle systems.
These vehicle systems include advanced driver assistance systems and vehicle guidance systems.
The development process proposed in the ISO~26262 standard is based upon multiple V-models, and defines activities and work products for each process step.
In many of these process steps, scenario based approaches can be applied to achieve the defined work products for the development of automated driving functions.
To accomplish the work products of different process steps, scenarios have to focus on various aspects like a human understandable notation or a description via state variables.
This leads to contradictory requirements regarding the level of detail and way of notation for the representation of scenarios.
In this paper, the authors discuss requirements for the representation of scenarios in different process steps defined by the ISO~26262 standard, propose a consistent terminology based on prior publications for the identified levels of abstraction, and demonstrate how scenarios can be systematically evolved along the phases of the development process outlined in the ISO~26262 standard.

\end{abstract}

\section{Introduction}

Driver assistance systems and automated systems reaching SAE Levels 1 and 2 \cite{SAE2016} have already been introduced to the market. 
Level 3 (conditional automation) and 4 (high automation) systems are announced to follow (Audi traffic jam pilot \cite{noauthor_techday_nodate} or Waymo self driving cars \cite{hawkins_waymo_2017}).
A challenge for the introduction of higher levels of automation is to assure that these vehicle systems behave in a safe way.
For driver assistance systems, this proof is furnished by driving many test kilometers on test grounds and public roads.
However, for higher levels of automation a distance-based validation is not an economically acceptable solution \cite{wachenfeld_release_2016}. 

As an alternative to the distance-based validation we introduce a scenario-based approach.
The key idea is to purposefully vary and validate the operating scenarios of the automated vehicle.
Therefore, the systematic derivation of scenarios and further assumptions have to be documented along the development process to ensure a traceable scenario generation.

The ISO~26262 standard is a guideline for the development of safety-critical electric/electronic vehicle systems and thus provides a framework for the development of vehicle guidance systems under the aspect of functional safety.
According to the ISO~26262 standard, scenarios can be utilized to support the development process.
For instance, scenarios can help to derive requirements, to develop the necessary hardware and software components, and to prove the safety of these components in the test process. 
When creating test cases, scenarios are necessary for generating consistent input data for the test object in any case.
Nevertheless, these different applications of scenarios result in distinct requirements for scenario representation in each development phase of the ISO~26262 standard. 

This contribution proposes three abstraction levels for scenarios along a V-model-based development process.
In this way, scenarios can be identified on a high level of abstraction in the concept phase and be detailed and concretized along the development process. 
This allows a structured approach, starting from the item definition according to the ISO~26262 standard, followed by the hazard analysis and risk assessment (HARA), and ending up with the necessary test cases for safety verification and validation.
Thus, the authors suggest an extended definition of the term `scenario' based on the definition of Ulbrich~et~al.~\cite{ulbrich_definition_2015} and introduce the abstraction levels of functional, logical, and concrete scenarios. 
A German version of this paper has been published at a workshop on driver assistance systems \cite{SzenarienProzess2017}.

The paper is structured as follows:
Section~\ref{related_work} gives a short motivation based on selected related work regarding scenarios in the development process for automated driving functions, utilized levels of abstraction for scenarios, and existing definitions of the term scenario.
Section~\ref{process} derives and analyzes requirements for the representation and usage of scenarios in the development process of the ISO~26262 standard.
Afterwards, section~\ref{terminologie} defines three layers of abstraction for scenarios and shows how these scenario representations can be converted into each other along the development process.
Finally, section~\ref{conclusion} gives a short conclusion.

\section{Related Work}
\label{related_work}
Ulbrich~et~al.~\cite{ulbrich_definition_2015} analyze the term \emph{scenario} across multiple disciplines and propose a consistent definition for the domain of automated vehicles.
In this paper, the authors use the term scenario referring to the definition of Ulbrich~et~al.~\cite{ulbrich_definition_2015}.

Go~and~Carroll~\cite{go_blind_2004} point out that scenarios have a different use across various disciplines, but the elements utilized to describe a scenario are similar in all cases.
Thereby, scenarios can be described in several levels of detail and different forms of notation.
Scenarios may be expressed in formal, semi-formal, or informal notation \cite{go_blind_2004}.
This distinction hints at multiple levels of abstraction of scenarios along the development process for automated vehicles.

Bergenhem~et~al.~\cite{bergenhem_how_2015} point out that complete requirements for vehicle guidance systems\footnote{To the authors' opinion, it is impossible to generate a complete set of requirements for higher levels of automation.} can only be achieved by a consistent, traceable, and verifiable process of requirements engineering in accordance with the V-model. 

Several publications suggest approaches which utilize scenarios to generate work products along the development process for automated vehicles.
Bagschik~et~al.~\cite{bagschik_identification_2016} develop a procedure for the generation of potentially hazardous scenarios within the process step of a hazard analysis and risk assessment, as suggested by the ISO~26262 standard.
This procedure utilizes an abstract description of the traffic participants and the scenery in natural language.
All possible combinations of scenario elements are analyzed incorporating descriptions of functional failures in a limited use case of an SAE Level 4 \cite{SAE2016} vehicle guidance system within the scope of the project \textit{Unmanned Protective Vehicle for Highway Hard Shoulder Road Works} (aFAS\footnote{This abbreviation is derived from the German project name.}) \cite{stolte_towards_2015}. 

Schuldt~et~al.~\cite{schuldt2011} motivate a scenario-based test process and present a systematic test case generation by use of a 4-layer-model.

Bach~et~al.~\cite{bach_model_2016} propose a model-based scenario representation with spatial and temporal relations as a general scenario notation along the development process of the ISO~26262 standard.
This scenario representation is implemented prototypically for scenarios of an ACC-system on motorways and the results are presented.

The mentioned publications utilize scenarios with different levels of abstraction for the functional and safety development of vehicle guidance systems.
The term `scenario' has not been defined uniformly, which makes it difficult to achieve a consistent understanding regarding the role of scenarios in the development process.
For this reason, the authors will derive and analyze requirements on scenarios in the following part.
%
%
\section{Scenario-based Design and Test Process referring to the ISO 26262 standard}
\label{process}
\begin{figure*}
	\centering
	\includegraphics[width=\textwidth]{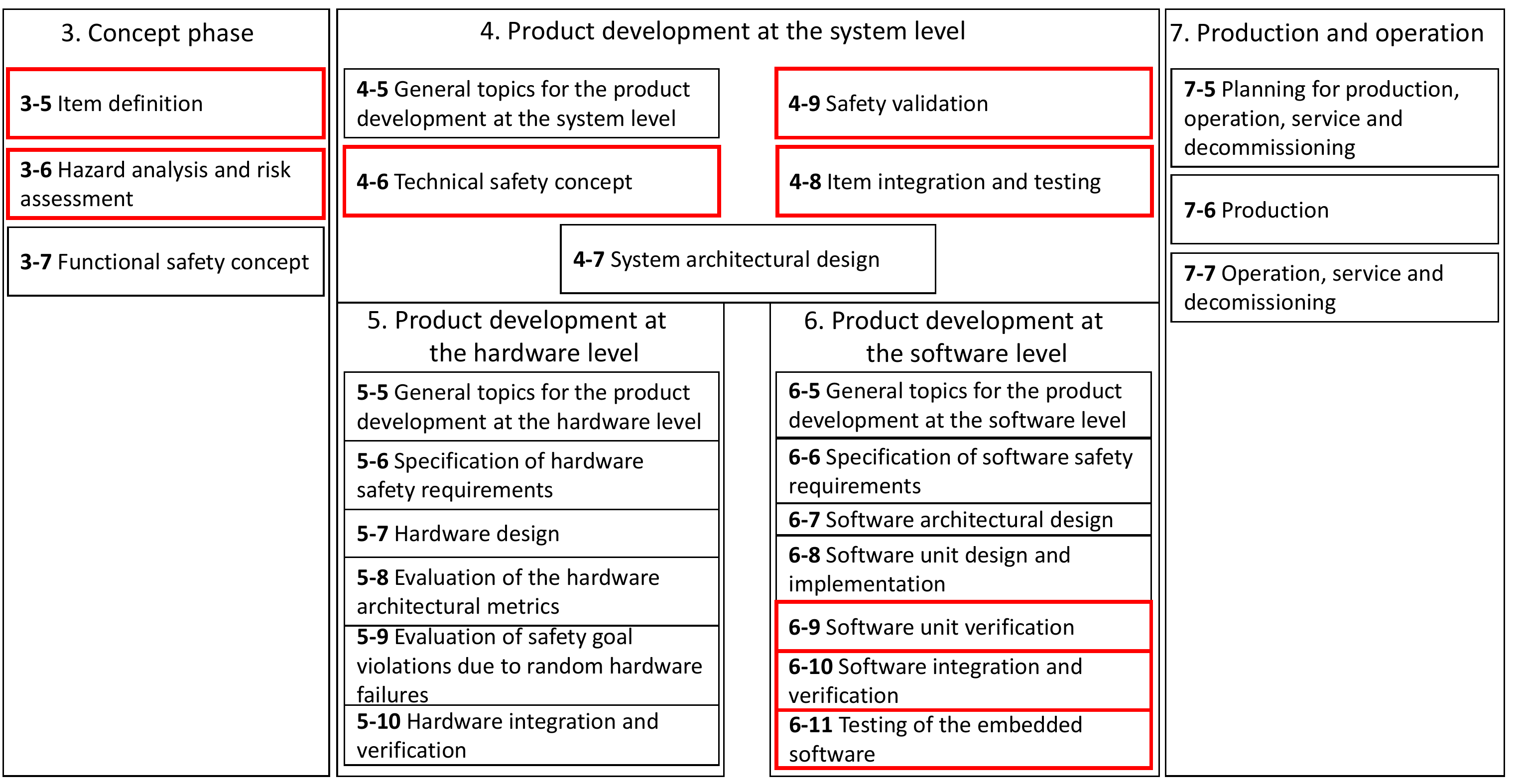}
	\caption{Overview of the development process proposed in the ISO~26262 standard. Process steps highlighted in red may utilize scenarios to generate the work products.}
	\label{fig:ISO-Prozess}
\end{figure*}

The ISO~26262 standard from 2016 \cite{ISO_26262_2011} represents the state of the art for developing vehicle guidance systems with regard to functional safety\footnote{The overall system development for vehicle guidance systems includes additional parallel development processes, which cover other aspects like function development.}.
An overview of the development process proposed in the ISO~26262 standard is shown in Fig.~\ref{fig:ISO-Prozess}. 
The process steps which may utilize scenarios to generate the demanded work products are highlighted in red.

Scenarios may support the whole development process of the ISO~26262 standard from the concept phase via the technical product development through to the system verification and validation. 
Hence, it is mandatory to define the requirements on scenarios resulting from the different process steps. 
These requirements allow a consistent definition of abstraction levels for the use of scenarios throughout the whole development lifecycle. 
The following sections refer to the work products of the development process defined by the ISO~26262 standard and derive requirements on scenarios for the highlighted process steps.

\subsection{Scenarios in the concept phase}
Prior to the technical development, the concept for the item under development is specified.
During the concept phase of the ISO~26262 standard (part 3) the item is defined, a hazard analysis and risk assessment is conducted, and a functional safety concept is developed.

The item definition shall include a description of the functional concept, system boundaries, the operational environment, the legal requirements, and the dependencies on other items.
Based on this information, possible operating scenarios can be derived. 
Reschka \cite{Reschka2016} proposes to identify safe driving states and specify the nominal behavior based on the operating scenarios.
The operating scenarios in this process step shall be described in an abstract level of detail and be represented in a human understandable way (textual description).

The next process step defined by the ISO~26262 standard which uses scenarios is the hazard analysis and risk assessment.
The hazard analysis and risk assessment consists of two steps: the situation analysis and the hazard identification, and the classification of hazardous events.
In the situation analysis, all operational situations\footnote{The authors point out that the term `operational situation' as it is used in the ISO~26262 standard should be declared as `operational scenario' according to Ulbrich~et~al.~\cite{ulbrich_definition_2015}.} and operating modes in which malfunctioning behavior will result in a hazardous event shall be described.
Whereby, malfunctioning behavior can be interpreted as deviation from the specified nominal behavior.
Afterwards, hazardous scenarios, which include a combination of operational scenarios and malfunctioning behavior, will be rated using the automotive safety integrity level (ASIL). 
The parameters for the ASIL classification are the exposure of the operational scenario, the possible severity, and the controllability of the hazardous scenario\footnote{The controllability of a scenario includes the controllability by the driver/passenger of the automated vehicle and the controllability by other traffic participants.}.
In order to determine these parameters, the description of hazardous scenarios has to include the stationary surroundings (scenery) and all traffic participants which may interact with the automated vehicle.

According to the actual state of the art, the analysis of hazardous scenarios is performed by experts. 
Hence, hazardous scenarios have to be formulated in natural language.
Depending on their area of expertise, human experts vary in the level of detail regarding the terms they use to describe a scenario.
Thus, a unified vocabulary for the functional perspective during the process step of the hazard analysis and risk assessment is necessary.
Furthermore, to ensure a common understanding among the experts, the terms within the vocabulary have to be organized in a semi-formal way.

Scenarios have to fulfill the following requirements to be utilized during the concept phase [C] of the ISO~26262 standard:
\begin{itemize}
	\item[C1] Human experts shall be able to formulate scenarios in the  field's terminology in natural language.
	\item[C2] Scenarios shall be represented in a semi-formal way.
\end{itemize}

\subsection{Scenarios in the system development phase}
Once the hazardous scenarios have been analyzed, a functional safety concept is developed.
To implement the functional concept, technical safety requirements have to be derived in process step 4-6.
As opposed to functional requirements, technical requirements outline criteria which can be physically quantified.
For example, the functional requirement to keep a safe driving distance to other traffic participants can be technically formulated by a distance in meters, which has to be satisfied.
Hence, every hazardous scenario has to be converted from the linguistic and semi-formal representation of the \textit{concept phase} to a representation via state values for the technical \textit{product development on system level} (4).
A list of those state variables is a precise description of a scenario, but, due to the high level of detail, not intuitively processable by human experts.
To reduce the quantity of scenarios, state values can be summarized in value ranges. 
Later on, those value ranges can be further detailed in valid/invalid ranges to define a set of safe and unsafe values respectively, or to model the system boundaries.
A detailed representation of scenarios ensures that the requirements on the item to be developed can be formulated in a verifiable way.
This is a necessary condition for the safety validation in process step 4-9 of the ISO~26262 standard.

All in all, scenarios have to fulfill the following requirements to be utilized during the system development phase [S] of the ISO~26262 standard:
\begin{itemize}
	\item[S1] Scenarios shall include the parameter ranges of the state values used for scenario representation.
	\item[S2] Scenarios shall provide a formal notation for the representation of the parameter ranges (for example a data format) to enable an automated processing.
\end{itemize}

\subsection{Scenarios for verification and validation}
During the test phase, it is examined whether the implemented system fulfills the requirements specified in the previous process steps.
For this verification, the tests have to be systematically planned, specified, executed, evaluated, and documented \cite[part 8, section 9.2]{ISO_26262_2011}.

Each test case specification has to include the following information independently from the test method \cite[part 8, section 9.4.2]{ISO_26262_2011}:
\begin{enumerate}
\item a unique identification
\item the reference to the work product to be verified
\item the preconditions and configurations\footnote{In the sense of a system variant.}
\item the environmental conditions
\item the input data including their time sequences
\item the expected behavior including acceptable variations
\end{enumerate}

A very challenging aspect of the test case generation is the specification of input data.
This data has to include time sequences of each parameter which is essentially affecting the behavior of the test object.
At the same time, due to highly connected systems, the input data may not contain any inconsistencies\footnote{Unintended inconsistencies are meant here. Fault injections can be utilized as a test method later on.}, but rather represent a consistent scenario.

Information regarding the operational environment of the system under verification as well as possible operating scenarios are already given in the item definition, which is specified during the concept phase of the development process according to the ISO~26262 standard.  
Based on this information, consistent input data can be derived for the specification of test cases.
The scenarios used in the item definition are expressed by language and formulated on an abstract level of detail. 
To utilize these abstract scenarios within the scope of a test case, the scenarios have to be specified in detail and concretized.

The detailed specification of scenarios can be performed within the scope of the \textit{specification of technical safety requirements} \cite[part 4, section 6]{ISO_26262_2011}.
The technical safety requirements describe how the item has to react to external stimuli which can affect the compliance with the safety goals.
In this way, the technical requirements also define for which parameter ranges the functionality of the system under development has to be ensured.
This parameter space has to be tested during the verification process and thus has to be taken into account for the test case generation.
In addition, the scenarios have to be converted to a formal representation during the step of specifying the scenarios in detail.
A formal representation is necessary, to ensure a reproducible test case execution later on.
The scenarios have to define all parameters required for test case execution via different test methods (like simulation or field tests).
Thus, in the step of specifying a scenario in detail, a conversion has to be conducted from an informal description based on organized terms to a formal description based on physical system state values.

To generate the input data included in a test case, discrete parameter values have to be chosen from the continuous parameter ranges of a specified scenario in a concretization step.
Schuldt \cite{Schuldt2017} proposes the use of equivalence classes, boundary value analysis, and combinatorial methods for identifying representative samples.
This approach provides a systematic generation of test cases, but lacks a method to determine a meaningful test coverage.
For determining a meaningful test coverage, the test concept, the scenario selection, and the necessary test methods have to be taken into account.
The scenarios, which are systematically derived during the concretization step and then formally described, represent consistent input data for the item under test.
Thus, the derived scenarios can be used in the scope of a test case for the verification of the implemented system.

All in all, scenarios have to fulfill the following requirements to be utilized during the testing phase [T] of the ISO~26262 standard:
\begin{itemize}
	\item[T1] Scenarios shall be modeled via concrete state values to ensure their reproducibility and to enable test methods to execute the scenario.
	\item[T2] Scenarios shall not include any inconsistencies.
	\item[T3] Scenarios shall be represented in an efficient machine readable way to ensure an automated test execution.
\end{itemize}

\subsection{Analysis of the derived requirements on scenarios}
Table \ref{tab:contradictoryRequirements} illustrates that the specified requirements are contradictory regarding the form of scenario description. 
On the one hand, requirement C1 states the demand for an abstract, linguistic scenario representation and, on the other hand, requirements S2 and T3 state the demand for an efficient, machine readable scenario representation.
Since linguistic representations are hard to process by machines and human beings are not able to read size efficient (mostly binary coded) data formats, there is a demand for different forms of scenario representations.

Similarly, requirements S1 and T2 demand different levels of detail for the scenario representation.
On the one hand, requirement S1 asks for a scenario representation via parameter ranges in the state space. 
This form of representation offers multiple degrees of freedom regarding the determination of concrete values to be tested.
On the other hand, requirement T2 asks for a representation that includes concrete parameter values. 
This form of representation is required for a reproducible test case execution.
Hence, machine readable scenarios have to support two different levels of detail.

\begin{table*}
	\centering
	\caption{Contradictory scenario requirements ($\perp$ marks contradiction)}
	\label{tab:contradictoryRequirements}
	\begin{tabularx}{\textwidth}{ X l X l X }
		\hline \\
		Concept phase & & System development phase & & Test phase \\
		\hline \\
		Human experts shall be able to formulate scenarios in the field's terminology in natural language. & \multirow{3}{4pt}{$\perp$} &  Scenarios shall include the parameter ranges of the state values used for scenario representation. & \multirow{3}{4pt}{$\perp$} &  Scenarios shall be modeled via concrete state values to ensure their reproducibility and to enable test methods to execute the scenario. \\
		\hline \\
	\end{tabularx}
\end{table*}

\section{Terminology for Scenarios along the Design and Test Process}
\label{terminologie}
\begin{figure*}
	\centering
	\includegraphics[width=1.0\textwidth]{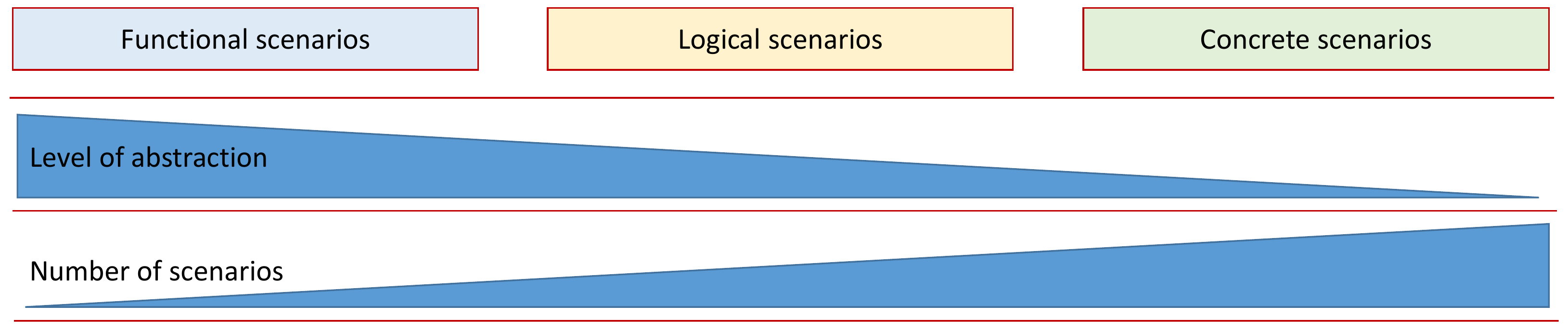}
	\caption{Levels of abstraction along the development process of the ISO~26262 standard}
	\label{fig:Abstraktionsebenen}
\end{figure*}

As stated in the previous section, the requirements on the type of scenario representations in the development process of the ISO~26262 standard are contradictory.
In the following section, the authors will suggest three abstraction levels for scenarios and show how these abstraction levels can be converted into each other along the development process.
Fig. \ref{fig:Abstraktionsebenen} illustrates the three levels of abstraction for scenarios: \textit{functional scenarios}, \textit{logical scenarios}, and \textit{concrete scenarios}.

\subsection{Functional scenarios}
Functional scenarios depict the most abstract level of scenario representations.
These scenarios may be used for the item definition and the hazard analysis and risk assessment during the concept phase of the ISO~26262 standard.
They are represented by language to ensure that human experts can easily understand existing scenarios, discuss them, and create new scenarios. 
The authors suggest the following definition:

\begin{quote}
\textit{
Functional scenarios include operating scenarios on a semantic level.
The entities of the domain and the relations of those entities are described via a linguistic scenario notation.
The scenarios are consistent.
The vocabulary used for the description of functional scenarios is specific for the use case and the domain and can feature different levels of detail.
}
\end{quote}

The representation of functional scenarios on a semantic level includes a linguistic and consistent description of entities and relations/interactions of those entities.
For the linguistic description a consistent vocabulary has to be defined.
This vocabulary includes terms for different entities (vehicle A, vehicle B) and phrases for the relations of those entities (vehicle A overtakes vehicle B).  

The required level of detail of functional scenarios depends on the actual development phase and the item under development.
Both aspects must be considered during the definition of the vocabulary.
For example, a highway pilot requires a vocabulary to describe the road geometry and topology, interactions with other traffic participants, and weather conditions. 
On the contrary, a parking garage pilot requires a vocabulary to describe the layout of the building whereas weather conditions may be irrelevant.
If a comprehensive vocabulary is used for the description of the entities and the relations of those entities, a large amount of scenarios can be derived from the vocabulary.
For a generation of consistent functional scenarios, all terms of the vocabulary have to be distinct.
Sources for terms that define the entities of a domain are, for example, actual standards and guidelines like road traffic regulations or the German standard for constructing motorways \cite{RAA}.

\begin{figure}
	\centering
	\includegraphics[width=0.9\columnwidth]{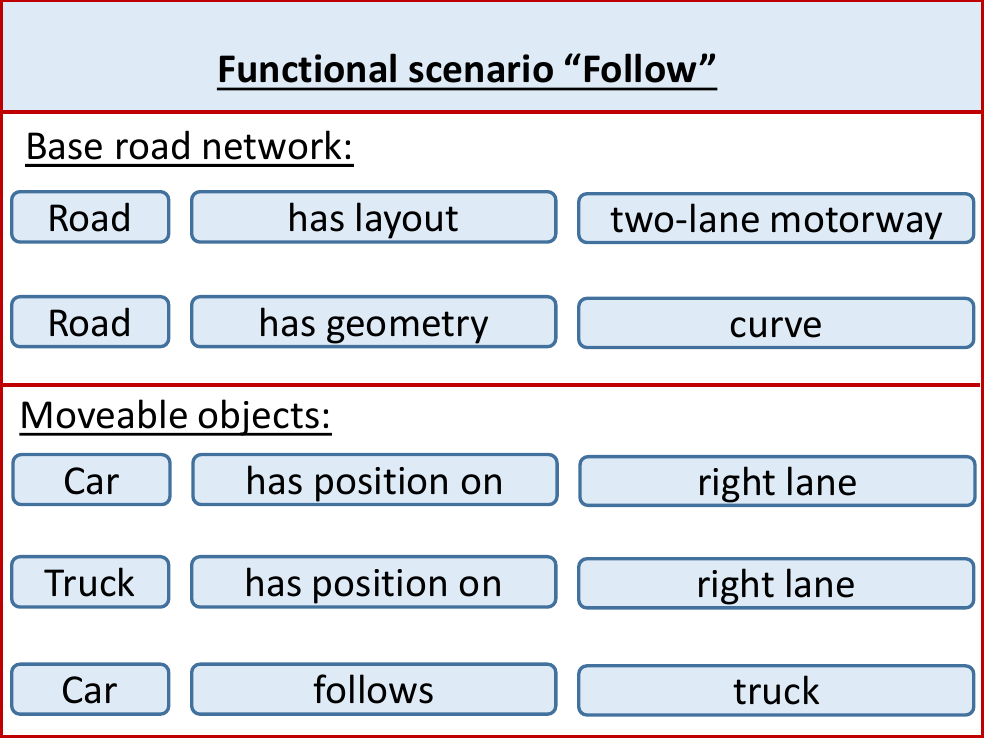}
	\caption{Example of a functional scenario. A car follows a truck on the right lane of a two-lane motorway in a curve.}
	\label{fig:functionalScenario}
\end{figure}

Fig. \ref{fig:functionalScenario} shows a functional scenario for a highway pilot on a two-lane motorway in a curve.
A car and a truck are driving on the right lane of the road, whereby the car follows the truck.

In this example, the road is described with a layout and a geometry.
Depending on the item's use case and domain, the vocabulary has to include additional terms to describe these characteristics like `three-lane motorway' for layout, and `straight' or `clothoid' for geometry.
The scenario can be varied by choosing other terms from the defined vocabulary.

\todo[inline]{Hier noch Moeglichkeit der automatisierten Erstellung einbringen? Gehoert mMn nicht zum Umfang des Papers. Wir haetten hier aber die Gelegenheit, unseren Forschungsbereich abzustecken.}


%
%

\subsection{Logical scenarios}
Logical scenarios depict a detailed representation of functional scenarios with the help of state space variables.
Those state space variables describe the entities and the relations of those entities.
Logical scenarios may be used to derive and represent requirements for the item during the system development phase.
For that purpose, logical scenarios describe the value ranges of the state space variables via a formal notation.
The authors suggest the following definition for logical scenarios:

\begin{quote}
\textit{
Logical scenarios include operating scenarios on a state space level.
Logical scenarios represent the entities and the relations of those entities with the help of parameter ranges in the state space.
The parameter ranges can optionally be specified with probability distributions.
Additionally, the relations of the parameter ranges can optionally be specified with the help of correlations or numeric conditions.
A logical scenario includes a formal notation of the scenario.
}
\end{quote}

The logical scenario description covers all elements necessary for the derivation of technical requirements needed to implement a system which solves these scenarios.
For a step-wise specification of scenarios in the development process of the ISO~26262 standard, logical scenarios have to be described via a formal notation in the state space, whereby parameters have to be defined via value ranges.
For a more detailed description of those parameter ranges, probability distributions (e.g., Gaussian distribution, Uniform distribution) can optionally be specified for each parameter range.
Additionally, relations of the parameter ranges can optionally be specified by numeric conditions (e.g., the speed of an overtaking vehicle has to be greater than the speed of the overtaken vehicle) or correlation functions (e.g., lane width correlates with curve radius).

\begin{figure}
	\centering
	\includegraphics[width=0.9\columnwidth]{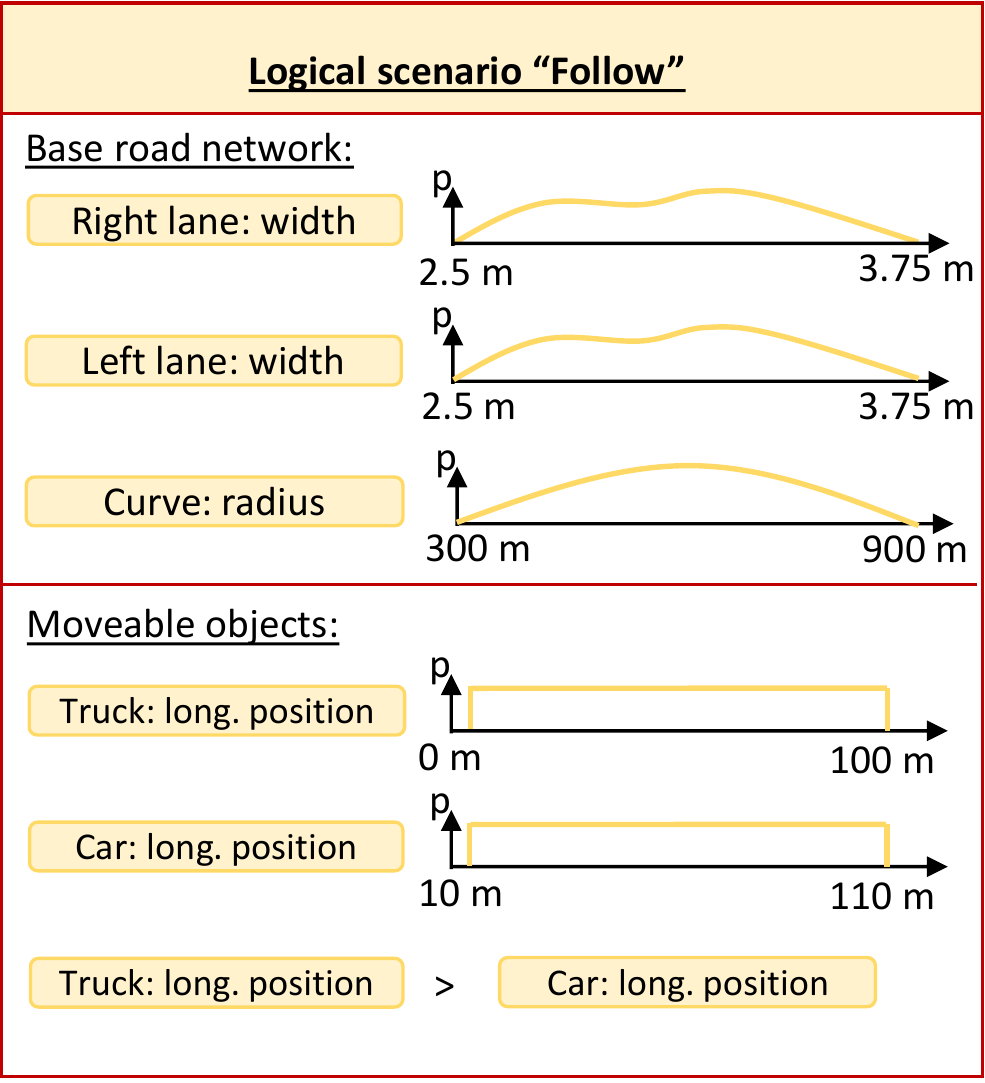}
	\caption{Example of a logical scenario. A car follows a truck on the right lane of a two-lane motorway in a curve.}
	\label{fig:logicalScenario}
\end{figure}

Fig. \ref{fig:logicalScenario} shows a logical scenario that has been derived from the functional scenario illustrated in Fig. \ref{fig:functionalScenario}.
Functional scenarios are converted to logical scenarios by a transformation from the linguistic representation into state space and specification of the scenario describing parameters.
Hence, every term from the vocabulary has to be assigned to parameters which describe this term.
In this example, both lanes are described via a lane width, the curve geometry is represented by a radius, and the vehicles are described by longitudinal positions along the lane.
Furthermore, the term `follows' demands that the longitudinal position of the truck is greater than the longitudinal position of the car.
To allow the example to be reflected in this paper, the authors have chosen a reduced set of parameters.
In reality, much more parameters will be necessary to describe a single term from the vocabulary.
For example, a truck can additionally be specified by its dimensions, weight, and engine power.

In addition, for every parameter from the example in Fig.~\ref{fig:logicalScenario} the value range and the probability distribution, with which the parameter occurs in reality, are specified.
This information helps to formulate technical requirements in the system development phase and provide a basis for a systematic generation of concrete scenarios in the testing phase.

%

\subsection{Concrete scenarios}
Concrete scenarios describe the entities and the relations for those entities using distinct parameters in the state space.
Every logical scenario can be converted to a concrete scenario by selection of a concrete value from a parameter range.
Concrete scenarios may be used as a basis for test case generation in the testing phase.
The authors suggest the following definition for concrete scenarios:

\begin{quote}
\textit{
Concrete scenarios distinctly depict operating scenarios on a state space level.
Concrete scenarios represent entities and the relations of those entities with the help of concrete values for each parameter in the state space.
}
\end{quote}

For each logical scenario with continuous value ranges any number of concrete scenarios can be derived.
For example, an infinite number of concrete scenarios can be achieved by choosing an infinitesimal sampling step width for each parameter.
An efficient concretization is accomplished by identification and combination of representative discrete values for each parameter.
Only concrete scenarios can directly be converted into test cases and executed with a vehicle guidance system.

\begin{figure}
	\centering
	\includegraphics[width=0.9\columnwidth]{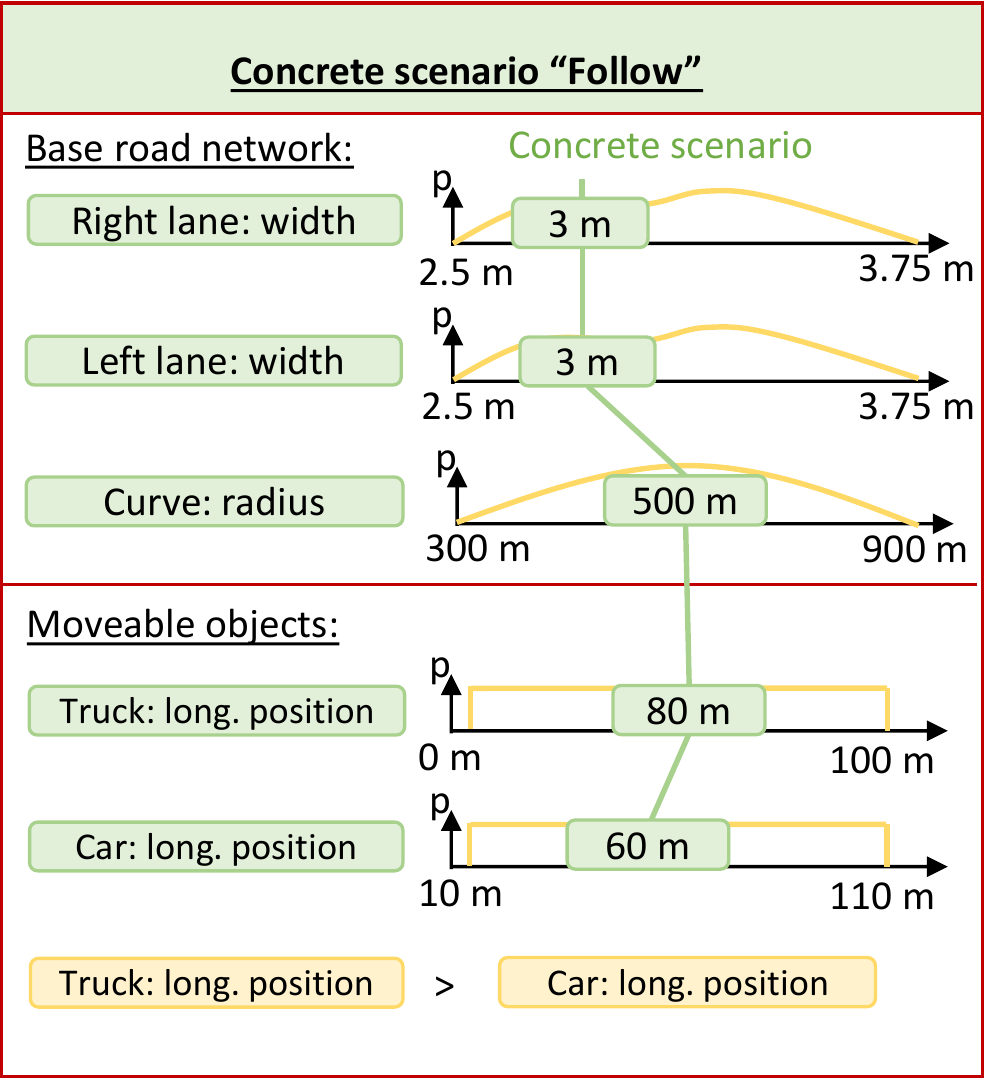}
	\caption{Example of a concrete scenario. A car follows a truck on the right lane of a two-lane motorway in a curve.}
	\label{fig:concreteScenario}
\end{figure}

Fig. \ref{fig:concreteScenario} shows a concrete scenario that has been derived from the logical scenario illustrated in Fig. \ref{fig:logicalScenario}.
For every parameter a concrete value within the defined value range has been chosen while the specified condition regarding the parameters has been satisfied.

To transform concrete scenarios into test cases, concrete scenarios have to be augmented by the expected behavior of the test object and the test infrastructure to be used as stated by Ulbrich~et~al.~\cite{ulbrich_definition_2015}.
The expected behavior can be derived from the functional operating scenarios, the logical scenarios, or the item definition. 

\section{Conclusion and outlook}
\label{conclusion}
In this paper, the authors analyzed the practicability of a scenario-based approach for the design of vehicle guidance systems following the development process of the ISO 26262 standard.
For this purpose, the process steps in which scenarios may be used to generate the work products of the respective process step have been identified.
Furthermore, requirements regarding the representation of scenarios have been defined and contradictions regarding the requirements resulting from different process steps have been shown.
On this basis, the authors suggested three levels of abstraction for scenarios in order to fulfill all requirements defined above.
Furthermore, a definition for each introduced level of abstraction has been given and it has been shown, how the levels of abstraction for scenarios can be used to generate work products for different process steps defined in the ISO~26262 standard.

In the future, new methods and tools are needed to generate functional scenarios and to convert these functional scenarios to concrete scenarios along the development process of the ISO~26262 standard. 
In addition to this contribution, there is a companion contribution submitted to the 2018 IEEE Intelligent Vehicles Symposium with an knowledge based approach for creating functional scenarios with a large variety.
Therefore, existing data formats for scenarios can be integrated into the suggested levels of abstraction.
Afterwards, new methods and tools for scenario specification and scenario concretization can be developed with respect to a test concept for automated vehicles.

\section{Acknowledgment}
We would like to thank the project members of the projects PEGASUS and aFAS funded by the German Federal Ministry for Economic Affairs and Energy for the productive discussions and the feedback on our approaches.
Our work is partially funded by the Volkswagen AG.

Additionally, we thank Andreas Reschka for his contributions to a first version of this paper.

\bibliographystyle{IEEEtran}
\bibliography{bib/bib}

\begin{thebibliography}{10}
\providecommand{\url}[1]{#1}
\csname url@samestyle\endcsname
\providecommand{\newblock}{\relax}
\providecommand{\bibinfo}[2]{#2}
\providecommand{\BIBentrySTDinterwordspacing}{\spaceskip=0pt\relax}
\providecommand{\BIBentryALTinterwordstretchfactor}{4}
\providecommand{\BIBentryALTinterwordspacing}{\spaceskip=\fontdimen2\font plus
\BIBentryALTinterwordstretchfactor\fontdimen3\font minus
  \fontdimen4\font\relax}
\providecommand{\BIBforeignlanguage}[2]{{%
\expandafter\ifx\csname l@#1\endcsname\relax
\typeout{** WARNING: IEEEtran.bst: No hyphenation pattern has been}%
\typeout{** loaded for the language `#1'. Using the pattern for}%
\typeout{** the default language instead.}%
\else
\language=\csname l@#1\endcsname
\fi
#2}}
\providecommand{\BIBdecl}{\relax}
\BIBdecl

\bibitem{SAE2016}
{Society of Automotive Engineers (SAE)}, ``J3016 - {Taxonomy and Definitions
  for Terms Related to On-Road Motor Vehicle Automated Driving Systems},''
  {Society of Automotive Engineers (SAE)}, 2016.

\bibitem{noauthor_techday_nodate}
\BIBentryALTinterwordspacing
``\BIBforeignlanguage{en}{{TechDay} piloted driving -- {The} traffic jam pilot
  in the new {Audi} {A}8},'' 2017, accessed: 01--15--2018. [Online]. Available:
  \url{https://www.audi-mediacenter.com/en/techday-piloted-driving-the-traffic-jam-pilot-in-the-new-audi-a8-9276}
\BIBentrySTDinterwordspacing

\bibitem{hawkins_waymo_2017}
\BIBentryALTinterwordspacing
``Waymo is first to put fully self-driving cars on {US} roads without a safety
  driver,'' 2017, accessed: 01--15--2018. [Online]. Available:
  \url{https://www.theverge.com/2017/11/7/16615290/waymo-self-driving-safety-driver-chandler-autonomous}
\BIBentrySTDinterwordspacing

\bibitem{wachenfeld_release_2016}
W.~Wachenfeld and H.~Winner, ``\BIBforeignlanguage{en}{The {Release} of
  {Autonomous} {Vehicles}},'' in \emph{\BIBforeignlanguage{en}{Autonomous
  {Driving}}}, M.~Maurer, J.~C. Gerdes, B.~Lenz, and H.~Winner, Eds.\hskip 1em
  plus 0.5em minus 0.4em\relax Berlin, Heidelberg, Germany: Springer Berlin
  Heidelberg, 2016, pp. 425--449.

\bibitem{ulbrich_definition_2015}
S.~Ulbrich, T.~Menzel, A.~Reschka, F.~Schuldt, and M.~Maurer, ``{Defining and
  Substantiating the Terms Scene, Situation, and Scenario for Automated
  Driving},'' in \emph{2015 {IEEE} 18th {International} {Conference} on
  {Intelligent} {Transportation} {Systems} ({ITSC})}, Las Palmas, Spain, 2015,
  pp. 982--988.

\bibitem{SzenarienProzess2017}
G.~Bagschik, T.~Menzel, A.~Reschka, and M.~Maurer, ``{Szenarien für
  Entwicklung, Absicherung und Test von automatisierten Fahrfunktionen -
  English title: Scenarios for Development, Test and Validation of Automated
  Vehicles},'' in \emph{11. {Workshop} {Fahrerassistenz und automatisiertes
  Fahren} {FAS} 2017}, Walting, Germany, 2017.

\bibitem{go_blind_2004}
K.~Go and J.~M. Carroll, ``The {Blind} {Men} and the {Elephant}: {Views} of
  {Scenario}-based {System} {Design},'' \emph{Interactions}, vol.~11, no.~6,
  pp. 44--53, 2004.

\bibitem{bergenhem_how_2015}
C.~Bergenhem, R.~Johansson, A.~S\"oderberg, J.~Nilsson, J.~Tryggvesson,
  M.~T\"orngren, and S.~Ursing, ``How to {Reach} {Complete} {Safety}
  {Requirement} {Refinement} for {Autonomous} {Vehicles},'' in \emph{{CARS}
  2015-{Critical} {Automotive} applications: {Robustness} \& {Safety}}, Paris,
  France, 2015.

\bibitem{bagschik_identification_2016}
G.~Bagschik, A.~Reschka, T.~Stolte, and M.~Maurer, ``Identification of
  {P}otential {H}azardous {E}vents for an {Unmanned} {Protective} {Vehicle},''
  in \emph{2016 {IEEE} {Intelligent} {Vehicles} {Symposium} ({IV})},
  Gothenburg, Sweden, 2016, pp. 691--697.

\bibitem{stolte_towards_2015}
T.~Stolte, A.~Reschka, G.~Bagschik, and M.~Maurer, ``Towards {Automated}
  {Driving}: {Unmanned} {Protective} {Vehicle} for {Highway} {Hard} {Shoulder}
  {Road} {Works},'' in \emph{2015 {IEEE} 18th {International} {Conference} on
  {Intelligent} {Transportation} {Systems} ({ITSC})}, Las Palmas, Spain, 2015,
  pp. 672--677.

\bibitem{schuldt2011}
F.~Schuldt, F.~Saust, B.~Lichte, M.~Maurer, and S.~Scholz, ``{Effiziente
  systematische Testgenerierung f\"ur Fahrerassistenzsysteme in virtuellen
  Umgebungen - English title: Efficient systematic test case generation for
  automated driving functions in virtual driving environments},'' in \emph{AAET
  - Automatisierungssysteme, Assistenzsysteme und eingebettete Systeme für
  Transportmittel}, Braunschweig, Germany, 2013, pp. 114 -- 134.

\bibitem{bach_model_2016}
J.~Bach, S.~Otten, and E.~Sax, ``Model based scenario specification for
  development and test of automated driving functions,'' in \emph{2016 {IEEE}
  {Intelligent} {Vehicles} {Symposium} ({IV})}, Gothenborg, Sweden, 2016, pp.
  1149--1155.

\bibitem{ISO_26262_2011}
ISO, \emph{26262 -- Road vehicles -- Functional Safety}, 2016.

\bibitem{Reschka2016}
A.~Reschka, ``{Fertigkeiten- und F\"ahigkeitengraphen als Grundlage f\"ur den
  sicheren Betrieb von automatisierten Fahrzeugen in st\"adtischer Umgebung -
  English title: Skills and ability graphs as basis for safe operation of
  automated vehicles in urban environments},'' Ph.D. dissertation, Technische
  Universit{\"a}t Braunschweig, 2017.

\bibitem{Schuldt2017}
F.~Schuldt, ``{Ein Beitrag f\"ur den methodischen Test von automatisierten
  Fahrfunktionen mit Hilfe von virtuellen Umgebungen - English title: Towards
  testing of automated driving functions in virtual driving environments},''
  Ph.D. dissertation, Technische Universit{\"a}t Braunschweig, 2017.

\bibitem{RAA}
\emph{Richtlinie f\"ur die Anlage von Autobahnen - English title: Guidelines
  for Constructing Motorways}, Forschungsgesellschaft f\"ur Stra{\ss}en und
  Verkehrswesen Std., 2009.

\end{thebibliography}

\end{document}